\documentclass[conference]{IEEEtran}
\IEEEoverridecommandlockouts
\usepackage{cite}
\usepackage{amsmath,amssymb,amsfonts}
\usepackage{algorithmic}
\usepackage{graphicx}
\usepackage{textcomp}
\usepackage{xcolor}
\def\BibTeX{{\rm B\kern-.05em{\sc i\kern-.025em b}\kern-.08em
    T\kern-.1667em\lower.7ex\hbox{E}\kern-.125emX}}

\begin{document}

\title{A Silicon Nitride Microring Based High-Speed, Tuning-Efficient, Electro-Refractive Modulator
}

\author{\IEEEauthorblockN{1\textsuperscript{st} Venkata Sai Praneeth Karempudi}
\IEEEauthorblockA{\textit{Department of ECE} \\
\textit{University of Kentucky}\\
Lexington, Kentucky, USA \\
kvspraneeth@uky.edu}
\and
\IEEEauthorblockN{2\textsuperscript{nd} Ishan G Thakkar}
\IEEEauthorblockA{\textit{Department of ECE} \\
\textit{University of Kentucky}\\
Lexington, Kentucky, USA \\
igthakkar@uky.edu}
\and
\IEEEauthorblockN{3\textsuperscript{rd} Jeffrey Todd Hastings}
\IEEEauthorblockA{\textit{Department of ECE} \\
\textit{University of Kentucky}\\
Lexington, Kentucky, USA \\
todd.hastings@uky.edu}

}

\maketitle

\begin{abstract}
The use of the Silicon-on-Insulator (SOI) platform has been prominent for realizing CMOS-compatible, high-performance photonic integrated circuits (PICs). But in recent years, the silicon-nitride-on-silicon-dioxide (SiN-on-SiO$_2$) platform has garnered increasing interest as an alternative to the SOI platform for realizing high-performance PICs. This is because of its several beneficial properties over the SOI platform, such as low optical losses, high thermo-optic stability, broader wavelength transparency range, and high tolerance to fabrication-process variations. However, SiN-on-SiO$_2$ based active devices such as modulators are scarce and lack in desired performance, due to the absence of free-carrier based activity in the SiN material and the complexity of integrating other active materials with SiN-on-SiO$_2$ platform. This shortcoming hinders the SiN-on-SiO$_2$ platform for realizing active PICs. To address this shortcoming, we demonstrate a SiN-on-SiO$_2$ microring resonator (MRR) based active modulator in this article. Our designed MRR modulator employs an Indium-Tin-Oxide (ITO)-SiN-ITO thin-film stack, in which the ITO thin films act as the upper and lower claddings of the SiN MRR. The ITO-SiN-ITO thin-film stack leverages the free-carrier assisted, high-amplitude refractive index change in the ITO films to effect a large electro-refractive optical modulation in the device. Based on the electrostatic, transient, and finite difference time domain (FDTD) simulations, conducted using photonics foundry-validated tools, we show that our modulator achieves 280 pm/V resonance modulation efficiency, 67.8 GHz 3-dB modulation bandwidth, $\sim$19 nm free-spectral range (FSR), $\sim$0.23 dB insertion loss, and 10.31 dB extinction ratio for optical on-off-keying (OOK) modulation at 30 Gb/s. 
\end{abstract}


\section{Introduction}
Driven by the rise of CMOS-compatible processes for fabricating photonic devices, photonic integrated circuits (PICs) are inexorably moving from the domain of long-distance communications to chip-to-chip and even on-chip applications. It is common for the PICs to incorporate optical modulators, to enable efficient manipulation of optical signals, which is a necessity for the operation of active PICs. Recent advances in the CMOS-compatible silicon-on-insulator (SOI) photonic platform has fundamentally improved the applicability of SOI PICs \cite{chrostowski2014}, \cite{bogaerts2020}, \cite{harris2018}. But in the last few years, the silicon-nitride-on-silicon-dioxide (SiN-on-SiO$_2$) platform has gained tremendous attention for realizing PICs. This is because the SiN-on-SiO$_2$ platform has several advantages over the SOI platform. Compared to silicon (Si), the SiN material has a much broader wavelength transparency range (500nm-3700nm), lower refractive index and smaller thermo-optic coefficient \cite{wilmart2019}. The lower refractive index of SiN means that SiN offers smaller index contrast with SiO$_2$ compared to Si. This in turn makes the SiN-on-SiO$_2$ based monomode passive devices (e.g., waveguides, microring resonators (MRRs)) less susceptible to \textit{(i)} propagation losses due to the decreased sensitivity to edge roughness \cite{bauters2011ultra}, and \textit{(ii)} aberrations in the realized device dimensions caused due to fabrication-process variations \cite{wilmart2019}. In addition, the smaller thermo-optic coefficient of SiN makes it possible to design nearly athermal photonic devices using SiN \cite{gao2017silicon}. Moreover, SiN devices and circuits exhibit increased efficiency of nonlinear parametric processes compared to Si \cite{levy2011integrated}.

Despite these favorable properties of the SiN-on-SiO$_2$ platform, SiN-on-SiO$_2$ based active devices such as modulators (e.g., \cite{phare2015graphene,alexander2018nanophotonic,jin2018piezoelectrically,ahmed2019high,hermans2019integrated}) are scarce and lack in modulation bandwidth, modulation efficiency and free spectral range (FSR) \cite{alexander2018nanophotonic}. This is because of the lack of the free-carriers based activity in the SiN material and the general difficulty of incorporating other active materials with the SiN-on-SiO$_2$ platform. This in turn limits the use of the SiN-on-SiO$_2$ platform for realizing only passive PICs. To overcome this shortcoming, there is impetus to heterogeneously integrate active photonic materials and devices with SiN-on-SiO$_2$ passive devices \cite{goyvaerts2021}. When such efforts of integrating electro-optically active materials with the SiN-on-SiO$_2$ platform come to fruition, it will be possible to design extremely high-performance and energy-efficient SiN-on-SiO$_2$ based active and passive PICs.

Different from such prior efforts, in this article, we demonstrate \textit{for the first time} the use of the high-amplitude electro-refractive activity of Indium-Tin-Oxide (ITO) thin films to realize a SiN-on-SiO$_2$ based optical on-off-keying (OOK) modulator. We show, based on the electrostatic, transient, and finite difference time domain (FDTD) simulations conducted using the photonics foundry-validated tools from Lumerical/Ansys, that our modulator achieves 280 pm/V resonance modulation efficiency, 67.8 GHz 3-dB modulation bandwidth, $\sim$19 nm free-spectral range (FSR), $\sim$0.23 dB insertion loss, and 10.31 dB extinction ratio for optical OOK modulation at 30 Gb/s. \textit{Based on the obtained simulation results, we advocate that our modulator can achieve better performance compared to the existing SiN modulators and several state-of-the-art Si and Lithium Niobate (LN) modulators from prior work}.

\section{Related Work and Motivation}
A plethora of Si, Lithium Niobate (LN) and SiN based integrated optical modulator designs have been formulated in prior work. But among these modulator designs, MRR based modulators have gained widespread attention due to their high wavelength selectivity, compact size, and compatibility for cascaded dense wavelength division multiplexing (DWDM). Here, we briefly review some relevant Si, LN and SiN MRR modulators from prior work.

\subsection{Silicon (Si) Based Modulators}
Over the last two decades, Si has emerged as the prominent material for fabricating PICs, mainly because the cost effectiveness of reusing the established CMOS manufacturing infrastructure promotes the use of Si for building complex PICs. Si material also exhibits high thermo-optic and electro-optic (free-carriers-induced) sensitivity, which enables the realization of optical modulators directly in Si substrate without requiring any auxiliary active materials. Si optical modulators based on free-carriers-induced plasma dispersion and absorption effects have become particularly more popular because of their low-power and high-speed operation. Although several designs of Si-based modulators have been reported, the most commonly adopted designs employ microring resonators (MRRs) (e.g., \cite{xu2005,xiao2012,moazeni2017,stojanovic2018,li2019silicon,van2012low,pantouvaki201556gb,sun2018128,ban2019low,li2020112}). The recent work \cite{li2019silicon} has also demonstrated the use of an electrically active stack of Si-SiO$_2$-ITO layers to substantially increase the electro-optic modulation efficiency of a Si MRR based modulator. In this design, the light-guiding core layer (i.e., Si) critically contributes to the electro-optic activity in the modulator. \textit{In contrast}, our SiN-on-SiO$_2$ modulator presented in this paper employs \textit{for the first time} an ITO-SiN-ITO thin-film stack in which the ITO thin films act as the active upper and lower claddings of the SiN MRR based core of the modulator.

\subsection{Lithium Niobate (LN) Based Modulators}
Lithium Niobate (LN) has recently emerged as the promising material for designing high-performance electro-optic modulators because of its wide bandgap and large second-order electro-optic coefficient. Several LN modulators have been demonstrated so far in the literature (e.g., \cite{chen2014,chen2015,wang2018,wang2019,lee2011}). For instance, in \cite{wang2018} and \cite{wang2019}, thin-film LN-based electro-optic modulators have been demonstrated. Similarly, in \cite{chen2014},\cite{chen2015} and \cite{lee2011}, a hybrid Si-LN platform based MRR modulators have been presented. These LN modulators demonstrated in prior works, however, lack in modulation efficiency compared to the Si and SiN modulators from prior works.

\subsection{Silicon Nitride (SiN) Based Modulators}
Recently, silicon nitride (SiN) based PICs have gained tremendous attention due to their favorable properties compared to the traditional Si based PICs. As a result, several SiN-on-SiO$_2$ modulators have been demonstrated (e.g., \cite{phare2015graphene,alexander2018nanophotonic,jin2018piezoelectrically,ahmed2019high,hermans2019integrated}). In \cite{phare2015graphene}, a graphene integrated electro-optic SiN MRR modulator has been reported. In \cite{ahmed2019high}, a hybrid SiN-LN platform based racetrack resonator modulator has been presented. Similarly, SiN modulators based on lead zirconate titanate and zinc oxide/zinc sulphide as active materials are demonstrated in \cite{alexander2018nanophotonic} and \cite{hermans2019integrated}. In \cite{jin2018piezoelectrically}, a SiN modulator that achieves tuning via photo-elastic effect has been demonstrated. Compared to these modulator designs from prior work, we present a different, ITO-based electro-refractive SiN-on-SiO$_2$ modulator that achieves relatively better modulation bandwidth, modulation efficiency, and FSR.

\section{Design of Our SiN-on-SiO$_2$ Modulator}
In this section, firstly we describe the structure and operating principle of our modulator design. Then, we discuss the characterization results for our modulator that we have obtained through photonics foundry-validated simulations. We also compare our modulator with several Si, LN and SiN based MRR modulators from prior work, in terms of modulation bandwidth, modulation efficiency, and FSR.

\begin{figure}
    \centering
    \includegraphics[scale = 0.25]{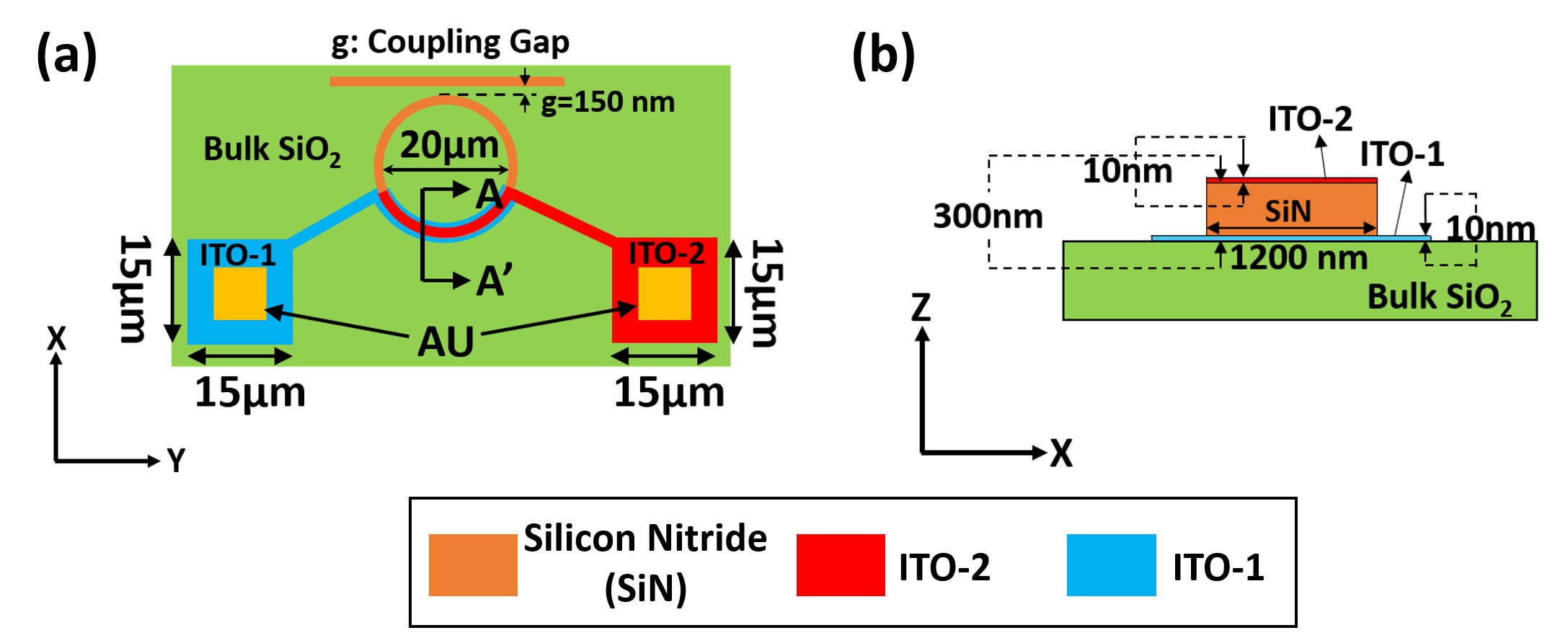}
    \caption{(a) Top view, (b) Cross-sectional view (along AA') of our SiN-on-SiO$_2$ MRR modulator.}
    \label{Fig:1}
\end{figure}

\begin{figure}
    \includegraphics[scale = 0.3]{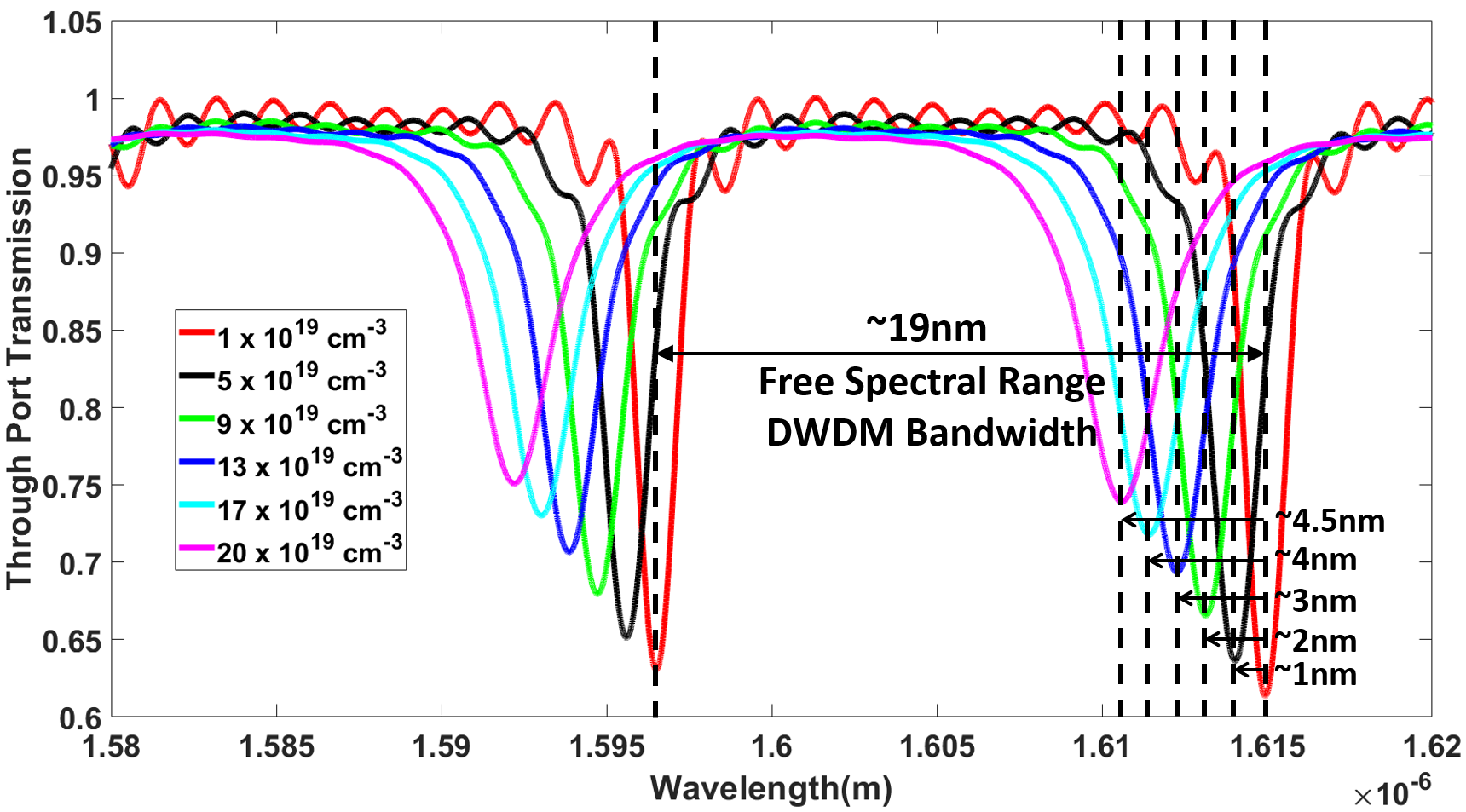}
    \caption{Transmission spectra of our modulator.}
    \label{Fig:2}
\end{figure}

\begin{figure}
    \centering
    \includegraphics[scale = 0.132]{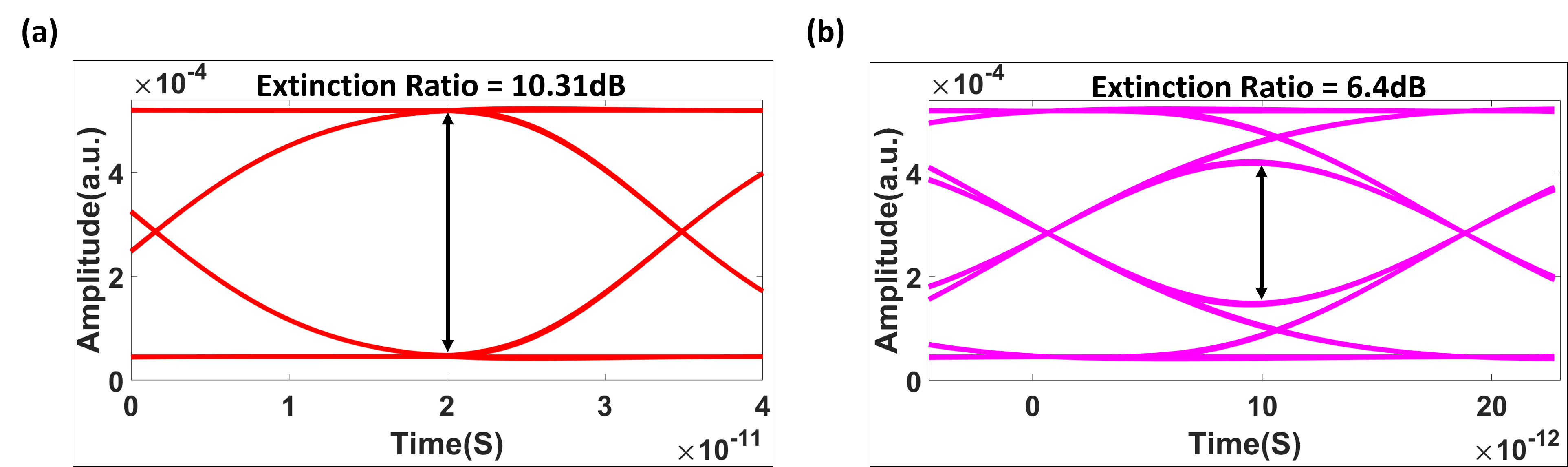}
    \caption{Optical eye diagrams for (a) 30 Gb/s and (b) 55 Gb/s OOK inputs to our modulator.}
    \label{Fig:3}
\end{figure}

\subsection{Structure and Operating Principle}
Fig. \ref{Fig:1}(a) and Fig. \ref{Fig:1}(b), respectively, show the top-view and cross-sectional schematics of our SiN-on-SiO$_2$ MRR modulator. The active region in the upper and lower claddings of the modulator consists of indium tin oxide (ITO) thin films with silicon nitride material (SiN) in between (creating an ITO-SiN-ITO thin-film stack). From Fig. \ref{Fig:1}(b), we have a 300 nm thick SiN-based MRR waveguide between two 10 nm thick ITO films. Upon applying voltage across the ITO-SiN-ITO stack (through the Au pads shown in Fig. \ref{Fig:1}(a)), free carriers accumulate in the ITO films at the ITO-SiN interfaces for up to ~5 nm depth in the ITO films \cite{chrostowski2014}, making these accumulation regions in the ITO films high-carrier-density active regions. This is due to the free-carriers-assisted, large-amplitude modulation in the permittivity and refractive index of the ITO material previously reported in \cite{chrostowski2014}. We evaluate this free-carriers based index modulation in the ITO films using the Drude-Lorentz model from \cite{ma2015indium}. It can be inferred from the Drude-Lorentz model that as the carrier concentration in the ITO accumulation regions increases, the refractive index of the ITO films decreases. Our modulator design from Fig. \ref{Fig:1} leverages this electro-refractive phenomenon in ITO. The free-carriers-induced decrease in the refractive index of the ITO thin films decreases the effective refractive index of the SiN-on-SiO$_2$ modulator, causing a blue shift in its resonance wavelength that in turn causes a transmission modulation in the MRR modulator. The electro-refractive activity of our SiN-on-SiO$_2$ MRR modulator is confined only in the ITO-based claddings. This is different from the Si-SiO$_2$-ITO capacitor based MRR modulator from \cite{li2019silicon}, which has the electro-refractive activity in both its Si-based MRR core and SiO$_2$-ITO based cladding.

\begin{figure}
    \centering
    \includegraphics[scale = 0.25]{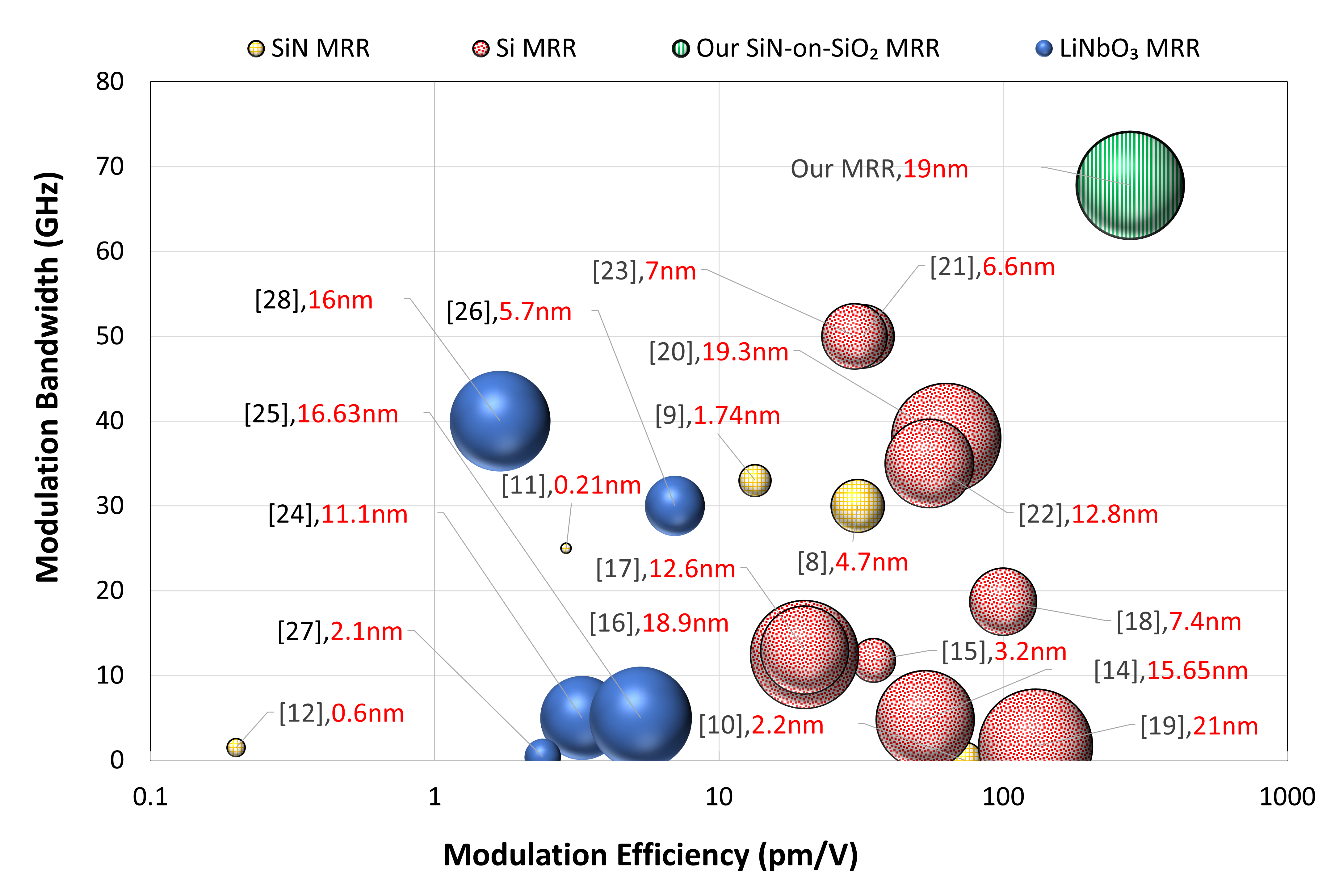}
    \caption{Modulation bandwidth, modulation efficiency and  FSR (shown as the size of the bubbles and red data labels) of various Si, LN (LiNbO$_3$) and SiN MRR modulators from prior work, compared with our SiN-on-SiO$_2$ MRR modulator.}
    \label{Fig:4}
\end{figure}

\subsection{Simulations based Characterization}
We performed electrostatic simulations of our ITO-SiN-ITO thin-film stack based SiN-on-SiO$_2$ modulator in the CHARGE tool of DEVICE suite from Lumerical \cite{lumericalwebsite}, to evaluate the required voltage levels across the Au pads (Fig. \ref{Fig:1}(a)) for achieving various free-carrier concentrations in the ITO films. Then, based on the Drude-Lorentz dispersion model from \cite{ma2015indium}, we extracted the corresponding ITO index change values for various free-carrier concentrations. These results are listed in Table \ref{Table:1}. Using these index values from Table \ref{Table:1}, we modeled our MRR modulator in the MODE tool from Lumerical \cite{lumericalwebsite} for finite-difference-time-domain (FDTD) and finite-difference eigenmode (FDE) analysis. For this analysis, we used the Kischkat model \cite{kischkat2012} of stoichiometric silicon nitride to model the MRR device. From this analysis, we extracted the effective index change and transmission spectra of our modulator (shown in Table \ref{Table:1} and Fig. \ref{Fig:2} respectively) at various applied voltages for the operation around 1.615 $\mu$m wavelength (L-band). From Fig. \ref{Fig:2}, our modulator achieves $\sim$4.5 nm resonance shift upon applying 17 V across the thin-film stack, which renders the resonance tuning (modulation) efficiency of $\sim$280 pm/V. This is crucially significant as our MRR modulator has relatively very low overlap between the optical mode and free-carrier perturbation (only $\sim$10\% of the guided optical mode overlaps with the ITO-based claddings) compared to the state-of-the-art ITO-based modulators (e.g., \cite{li2019silicon}). Further, from the simulated spectra in Fig. \ref{Fig:2}, we evaluate the FSR of our modulator to be $\sim$19 nm. Further, based on our device simulations using the Lumerical MODE tool, we evaluated the insertion loss and loaded Q-factor of our modulator to be $\sim$0.23 dB and $\sim$2300 respectively. We also evaluated the capacitance density of the ITO thin-films covering the MRR rim (using the Lumerical CHARGE tool) to be $\sim$0.18 fF$/\mu$m$^2$ for the 300 nm thick SiN layer, yielding the modulation bandwidth (3-dB RC bandwidth) of $\sim$67.8 GHz for the modulator. We also modeled our modulator in Lumerical INTERCONNECT, to  simulate optical eye diagrams for the modulator at 30 Gb/s and 55 Gb/s operating bitrates (Fig. \ref{Fig:3}). As evident (Fig. \ref{Fig:3}(b)), our modulator can achieve 10.31 dB extinction ratio for OOK modulation at 30 Gb/s bitrate.

\begin{table}[]
\caption{Free-carrier concentration (N), real index (Re($\eta_{ITO}$)), and imaginary index (Im($\eta_{ITO}$)) for the ITO accumulation layer in our modulator. The real and imaginary effective index (Re($\eta_{eff}$), Im($\eta_{eff}$)), operating voltage (V), and induced resonance shift ($\Delta\lambda_{r}$) for our modulator.}
\begin{tabular}{|c|c|c|c|c|c|c|}
\hline
\textbf{\begin{tabular}[c]{@{}c@{}}N\\($cm^{-3}$)\end{tabular}} & \textbf{\begin{tabular}[c]{@{}c@{}}Re\\
($\eta_{ITO}$)\end{tabular}} & \textbf{\begin{tabular}[c]{@{}c@{}}Im\\
($\eta_{ITO}$)\end{tabular}} & \textbf{\begin{tabular}[c]{@{}c@{}}Re\\
($\eta_{eff}$)\end{tabular}} & \textbf{\begin{tabular}[c]{@{}c@{}}Im\\
($\eta_{eff}$)\end{tabular}} & \textbf{V} & \textbf{\begin{tabular}[c]{@{}c@{}}$\Delta\lambda_{r}$\\   (pm)\end{tabular}} \\ \hline
1×10$^{19}$                                                         & 1.9556                                                        & 0.0100                                                         & 1.9973                                                        & 2.651e-5                                                       & 0          & 0                                                            \\ \hline
5×10$^{19}$                                                          & 1.9111                                                        & 0.0403                                                         & 1.99434                                                        & 2.6581e-5                                                      & 3.4        & 991                                                          \\ \hline
9×10$^{19}$                                                          & 1.8667                                                        & 0.0896                                                         & 1.99138                                                        & 2.6587e-5                                                      & 6.8        & 1890                                                         \\ \hline
13×10$^{19}$                                                         & 1.8222                                                        & 0.1289                                                         & 1.98842                                                        & 2.6593e-5                                                      & 10.2       & 2970                                                         \\ \hline
17×10$^{19}$                                                         & 1.7778                                                        & 0.1582                                                         & 1.98546                                                        & 2.6598e-5                                                      & 13.6       & 3910                                                         \\ \hline
20×10$^{19}$                                                         & 1.7333                                                        & 0.1874                                                         & 1.9825                                                        & 2.6604e-5                                                      & 17.0       & 4470                                                         \\ \hline
\end{tabular}
\label{Table:1}
\end{table}

\subsection{Comparison and Discussion}
Fig. \ref{Fig:4} shows a comparison of our SiN-on-SiO$_2$ modulator with the best performing Si (ten; \cite{xu2005}-\cite{li2020112}), LN (five; \cite{chen2014}-\cite{lee2011}) and SiN (five; \cite{phare2015graphene}-\cite{hermans2019integrated}) MRR modulators from prior work, in terms of three key attributes, namely modulation efficiency, FSR, and modulation bandwidth. As evident from Fig. \ref{Fig:4}, our modulator achieves better performance compared to the exisiting SiN modulators and the state-of-the-art Si and LN modulators from prior works, which in turn promotes its use in DWDM-based high-performance PICs. Since our SiN-on-SiO$_2$ modulator achieves modulation bandwidth of $\sim$67.8 GHz, it can be easily operated at the bitrate of $>$15 Gb/s to enable ultra-high-speed (potentially beyond Tb/s) DWDM-based PICs while ensuring minimal power-penalty from crosstalk \cite{bahadori2016crosstalk} and self-heating \cite{seyedi201615}. In addition, our SiN-on-SiO$_2$ modulator also achieves a modulation efficiency of $\sim$280 pm/V. This in turn can enable dynamic operation of our modulator with energy-efficiency of $<$100 fJ/bit \cite{AlanWang2019}. Unfortunately, our modulator achieves relatively low loaded Q-factor of 2300. Nevertheless, we anticipate that the Q-factor can be increased to 5000-8000 by marginally trading the modulation bandwidth for better loss characteristics of the MRR cavity. Thus, future work should include an exhaustive search of design parameters, including the coupling gap, MRR waveguide width, MRR waveguide height, MRR radius, and the thicknesses of the ITO films, to minimize the coupling, bending and absorption losses in the MRR cavity without notably compromising the modulation efficiency. Having the loaded Q-factor of our modulator in the range of 5000-8000, while already having a greater than 12 nm FSR (Fig. \ref{Fig:4}), will enable balancing of the crosstalk penalty and modulation speed in our modulator, for high-performance DWDM based PICs \cite{bahadori2016crosstalk}. Moreover, although ITO is not available in the CMOS process flow, it can be deposited at relatively low temperatures (less than 300°C) on top of the back-end-of-line (BEOL) metal layers of CMOS chips, independent of the CMOS FEOL process. This makes our SiN-on-SiO$_2$ modulator an excellent choice for implementing optical interconnect PICs on silicon interposers, to enable ultra-high-bandwidth inter-chiplet communication in emerging multi-chiplet systems \cite{Popstar}. 

\textit{In summary}, we advocate that our SiN-on-SiO$_2$ modulator can achieve better performance compared to the SiN, Si and LN based MRR modulators from prior work. The obtained results corroborate our modulator's potential to consequently enable DWDM-based SiN-on-SiO$_2$ PICs that will offer highly scalable and energy-efficient solutions to a wide range of mature and emerging applications, including datacenter transceivers \cite{buckwalter2012monolithic}, high-performance computing \cite{rumley2015silicon}, signal processing \cite{khilo2012photonic}, optical computing \cite{Karempudi2021SenSys}, and artificial intelligence \cite{shiflett2020pixel}.

\section{Conclusion}
In recent years, the SiN-on-SiO$_2$ platform has attained tremendous attention for realizing PICs because it has several advantageous properties over the conventional SOI platform. Despite these advantages, the SiN-on-SiO$_2$ platform lacks high-performance active devices such as modulators. To address this drawback, we have demonstrated an ITO based SiN-on-SiO$_2$ MRR modulator, which consists of ITO thin films as the active upper and lower claddings of the SiN MRR core. These ITO-based active claddings of our modulator leverage the free-carrier assisted, high-amplitude refractive index change in them to effect a large electro-refractive optical modulation in the device. To evaluate the performance of our SiN-on-SiO$_2$ MRR modulator, we performed electrostatic, transient and finite difference time domain (FDTD) simulations using the foundry-validated Ansys/Lumerical tools. Based on these simulations, our modulator achieves superior performance with $\sim$280 pm/V modulation efficiency, 67.8 GHz 3-dB modulation bandwidth with $\sim$19nm FSR, $\sim$0.23 dB insertion loss and 10.31 dB extinction ratio for OOK modulation at 30 Gb/s. This excellent performance of our SiN-on-SiO$_2$ MRR modulator demonstrates its potential to enhance the performance and energy-efficiency of SiN-on-SiO$_2$ based PICs of the future.

\bibliographystyle{IEEEtran}
\bibliography{ref}

\end{document}